\documentclass[intlimits,twoside,a4paper]{article}

\usepackage{amsmath,amssymb}
\usepackage{graphicx}
\usepackage{siunitx}
\usepackage[T2A]{fontenc}
\usepackage[cp1251]{inputenc}

\usepackage[eqsecnum]{cmpj2}

\articletype{Proceedings Paper}

\issue{2013}{16}{3}{31701}
\doinumber{10.5488/CMP.16.31701}

\title[Dielectric  studies on  strontium-barium niobates]%
{Ferroelectric and dielectric characterization studies on relaxor- and
ferroelectric-like strontium-barium niobates}

\author[K. Matyjasek \textsl{et al.}]{K. Matyjasek\refaddr{label1},
       J. Dec\refaddr{label2},
		  S. Miga\refaddr{label2},
		  T. \L{}ukasiewicz\refaddr{label3}	}
\addresses{
\addr{label1} Institute of Physics, Faculty of Mechanical Engineering and Mechatronics, West Pomeranian University of Technology, Al. Piastow 48, 70--310 Szczecin, Poland
\addr{label2} Institute of Materials Science, University of Silesia, 12 Bankowa St., 40--007 Katowice, Poland
\addr{label3} Institute of Electronic Materials Technology, 133 W\'{o}lczy\'{n}ska St. 01--919 Warsaw, Poland}

\authorcopyright{K. Matyjasek, J. Dec, S. Miga, T. \L{}ukasiewicz, 2013}
\date{Received October 3, 2012, in final form November 21, 2012}

\begin{document}
\maketitle

\begin{abstract}
Ferroelectric domain structure  evolution induced by an external electric field was investigated by means of nematic liquid crystal (NLC) method in two strontium-barium niobate single crystals of nominal composition: Sr$_{0.70}$Ba$_{0.30}$Nb$_{2}$O$_{6}$ (SBN:70~--- relaxor) and  Sr$_{0.26}$Ba$_{0.74}$Nb$_{2}$O$_{6}$ (SBN:26~--- ferroelectric). Our results provide evidence that the broad phase transition and frequency dispersion that are exhibited in SBN:70 crystal have a strong link to the configuration of ferroelectric microdomains. The large leakage current revealed in SBN:26 may compensate internal charges acting as pinning centers for domain walls, which gives rise to a less restricted  domain growth similar to that observed in classical ferroelectrics. Microscale studies of a switching process in conjunction with electrical measurements allowed us to establish a relationship between local properties of the domain dynamics and macroscopic response i.e., polarization hysteresis loop and dielectric properties.

\pacs 77.80.Dj, 77.80.Fm, 77.80.Jk, 77.84.Dy, 77.22.Ej
\end{abstract}

\section{Introduction}
Strontium-barium niobates, SBN (Sr$_{x}$Ba$_{1-x}$Nb$_{2}$O$_{6}$, $0.25 \leqslant x
\leqslant 0.75$), are ferroelectric crystals that have received a great interest in
their applications in optoelectronics \cite{Ramirez12}. The
controlled manipulation of the domain structure via external electric field
involves an important issue of potential applications, such as optical-frequency
conversion and high density optical data storage \cite{Soergel05}. Thus, efforts
are directed towards  producing SBN crystals with high quality ferroelectric
domain reversible structures. Polarization switching, which proceeds by
nucleation and growth of domains, has attracted much attention also from the
viewpoint of statistical physics related to solid-state transformation.

Only a few works focusing on SBN domain structure dynamics can be found in
literature. Mostly, high-resolution domain structure studies (nanoscale domains)
have been performed for a congruently melting composition
Sr$_{0.61}$Ba$_{0.39}$Nb$_{2}$O$_{6}$ (SBN:61) by piezoresponse force microscopy
(PFM) \cite{Shur11,Terabe02,Lehnen01,Shvartsman08,Liu09,Gainutdinov09}. The
microscale domain structure kinetics in SBN:61 was investigated using an
electro-optic imaging microscope \cite{Tian05,Shur08} and NLC method
\cite{Ivanov02,Matyjasek08,Matyjasek12}. The observed specific features of the
switching process were accounted for freezing or pinning domain walls connected
with disordered structure of SBN crystals.

SBN has an open tetragonal tungsten bronze structure, in which only five of six
available positions of Sr$^{2+}$  and Ba$^{2+}$ cations are occupied
\cite{Jamieson68}. The origin of relaxor behaviour in SBN can be attributed to
the development of a quenched random fields associated with the
composition/structural disorder \cite{Nattermann1990,Kleemann98}. The ferroelectric properties
of SBN system change with the composition \cite{Qu02,Lukasiewicz08}. To the best
of our knowledge, there is no report on the domain structure dynamics  in other
SBN compositions, in spite of its significant effect on the physical properties.
Temperature dependence of dielectric permittivity $\varepsilon(T)$, at various
frequencies has shown a gradual crossover from typical relaxor behaviour in
SBN:75 into classical ferroelectrics as observed in the case of SBN:40
\cite{Lukasiewicz08,Santos09}. It is reflected by a broad peak of
$\varepsilon(T)$ function, which was strongly dependent on frequency for SBN:75
crystal; while a sharper peak, with the position of the maximum of
$\varepsilon(T)$ function independent of frequency, was observed for SBN:40.

In this report, we have used  the NLC method to study the domain creation
process in static electric fields, in SBN:70 and SBN:26 compositions. The NLC
method involves the averaging over macroscopic scale and this enables us to correlate
the domain structure dynamics with macroscopic characterization techniques such
as electric displacement~-- electric field ($D-E$) hysteresis loop, switching
currents and dielectric permittivity measurements. Investigating the change from
relaxor (SBN:70) to classical ferroelectric behaviour (SBN:26) is thus helpful
for better understanding the physics of the relaxor ferroelectrics.

\section{Experimental details}

SBN crystals, as uniaxial ferroelectric materials, exhibit only 180$^{\circ}$
domains because the paraelectric phase has a tetragonal symmetry (4/mmm), and
the order parameter in the ferroelectric phase (4mm) occurs along [001]
direction \cite{Jamieson68}. Details concerning the crystal fabrication by
Czochralski method are given elsewhere \cite{Lukasiewicz08}. Two large single
crystals of nominal composition Sr$_{0.70}$Ba$_{0.30}$Nb$_2$O$_6$ (SBN:70) and
Sr$_{0.26}$Ba$_{0.74}$Nb$_2$O$_6$ (SBN:26) have been grown. Using an Inductively
Coupled Plasma~-- Optical Emission Spectroscopy (ICP--OES), a real composition of
the grown single crystals was determined to be \linebreak
Sr$_{0.70}$Ba$_{0.26}$Nb$_{2}$O$_{5.96}$ and
Sr$_{0.35}$Ba$_{0.69}$Nb$_{2}$O$_{6.04}$, respectively. The obtained  single crystals seem
to be slightly non-stoichiometric. Despite an obvious difference
between the nominal and real compositions, the samples are labelled throughout
the paper by their nominal stoichiometry. Platelet-shaped samples were cut
perpendicular to the [001] direction and polished to the optical quality.

The NLC mixture of p-methoxybenzylidene-p-n-butylaniline (MBBA) and
pethoxybenzylidene-p-n-butylaniline (EBBA) was used to observe optically indistinguishable 180$^{\circ}$ domain walls. The NLC method makes possible a
continuous observation of the domain pattern during polarization reversal in
an electric field if a cover glass coated with a conducting layer of tin oxide
is used.  The reversed regions, where reorientation of domains still occurs,
look somewhat darker than the ``grey'' surrounding domains, because
a certain electrohydrodynamic instability, particularly dynamic scattering,
takes place in these regions \cite{Tikhomirova79}. The ($D-E$) hysteresis loops were recorded
using a modified Sawyer-Tower circuit by using an $ac$ field of frequency 50~Hz
and a digital oscilloscope. Switching current transients were measured using a
wave-form function generator, a small standard resistor and digital
oscilloscope. The complex linear susceptibility $\chi=\chi'-\ri\chi''$ was
measured using a Solartron 1260 Impedance Analyzer together with a 1296
Dielectric Interface (SBN70) and an Agilent E4980A Precision LCR Meter (SBN26)
applying a weak probing $ac$ electric field of the order of 2~Vcm$^{-1}$. For
electric and dielectric studies, the SBN samples were prepared as plates with
dimensions $5\times5\times0.5$~mm$^3$ with Cu-Au electrodes evaporated
onto the principal (001) faces. Thin copper interface is to improve the
adhesion of gold. The temperature protocols required by experiment were managed
using a Lake Shore Model 340 temperature controller.

\section{Experimental results}
\subsection{Ferroelectric characteristics}

\begin{figure}[!tb]
\centerline{
\includegraphics[width=0.55\textwidth]{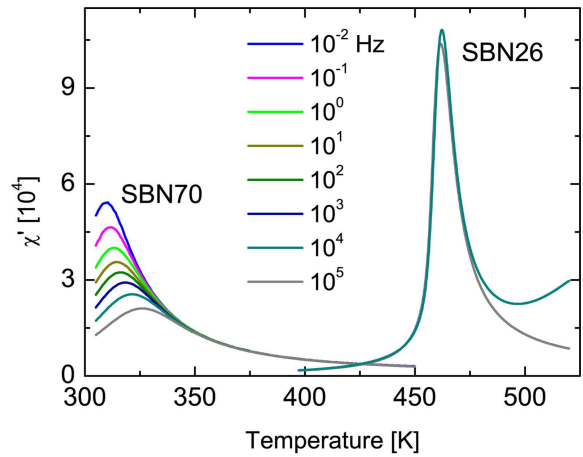}
}
\caption{(Color online) Temperature dependences of the real part of the electric susceptibility of SBN:70 and SBN:26 measured at different frequencies, $10^{-1} \leqslant f\leqslant  10^5$~Hz; due to an enhanced electric conductivity (leakage current) of SBN26 only data measured at two highest frequencies are reliable.} \label{fig1}
\end{figure}

Figure~\ref{fig1} presents the temperature dependences of real parts of electric
susceptibility of SBN:70 and SBN:26 measured at different frequencies. The
SBN:70 displays a typical relaxor behavior where the temperature position of the
broad maximum of susceptibility and its height strongly depend on the
frequency of the driving electric field. On the other hand, SBN:26 shows much
higher susceptibility peaks ($\chi'=115000$) taking place at a fixed
temperature, $T=461$~K. These maxima might be considered as a sign of a
phase transition between an ordered ferroelectric and paraelectric states.
Since the electric conductivity (leakage current) of SBN:26 increases
distinctly at higher temperatures, the measurements of susceptibility were
carried out only at higher frequencies of the probing field  ($10^4$ and
$10^5$~Hz). Under this condition the effect of the leakage current on the
dielectric response may be neglected.

\begin{figure}[!b]
\centerline{
\includegraphics[width=0.65\textwidth]{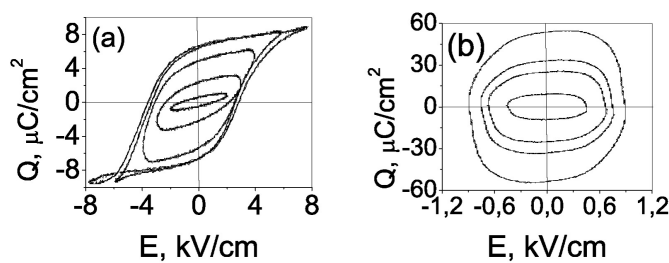}
}
\caption{The family of $D-E$ hysteresis loops obtained at room temperature by applying $ac$-field at a frequency of 50~Hz: (a)~--- SBN:70 and (b)~--- SBN:26 crystal.} \label{fig2}
\end{figure}

Figure~\ref{fig2} shows the room temperature  $D-E$ loops for SBN:70 (a) and SBN:26 (b)
compositions. These loops were obtained by applying a sinusoidal signal at a
frequency of 50~Hz. The polarization charges accumulated on the crystal sample
result from various effects of the crystal bulk, such as changes in the domain
structure and leakage currents. The $D-E$ loops of the SBN:26 sample, present an
oval-like profile owing to the large leakage current in the crystal sample. It
is difficult to determine the coercive field, $E_\mathrm{c}$ which increases steadily
with an increase of the electric field amplitude. These results suggest that
even at high electric fields, there are frozen regions that do not participate
in the switching polarization and/or that a large contribution to the
polarization is due to the sidewise domain wall motion, which is a very slow
mechanism.

\subsection{Domain switching in SBN:70 crystal}

\begin{figure}[!tb]
\centerline{
\includegraphics[width=0.9\textwidth]{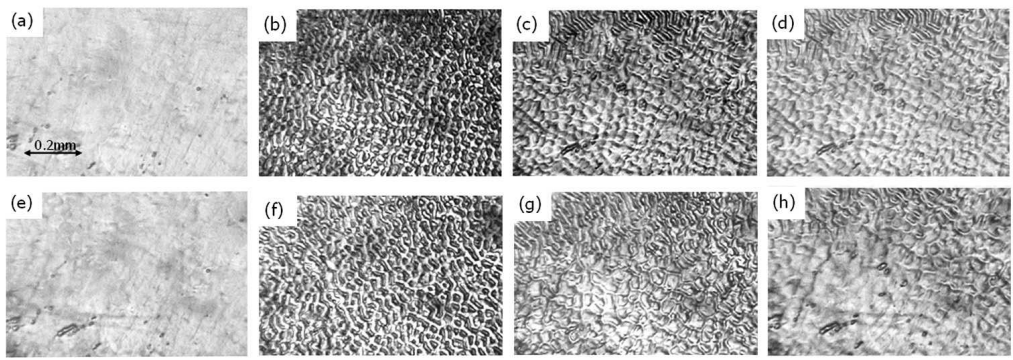}
}
\caption{Domain configurations developing at different stages of polarization reversal process, observed on the (001) plane of SBN:70 crystal sample, during switching in electric field from 1~kVcm$^{-1}$ to 1.5~kVcm$^{-1}$. (a)~--- initial single domain state. Time from the moment of applying $E=1$~kVcm$^{-1}$: (b) 0.4~s, (c) 0.8~s, (d) 2~s, (e) 5~s~--- the domain walls are rather difficult to distinguish because of their slow velocity, (f) distinct variations of the domain pattern are noticeable following the application of $E=1.3$~kVcm$^{-1}$, (g) the picture of delineated domain walls 4~s after electric field application of 1.3~kVcm$^{-1}$, (h) the switching process was completed after application of $E=1.5$~kVcm$^{-1}$.} \label{fig3}
\end{figure}

To investigate the domain dynamics, the crystal sample was
preliminary poled into a single domain state with a sufficiently high electric
field $E>E_\mathrm{c}$, and then an electric field of the opposite direction was applied to
the sample. The formation of domains is possible if the amplitude of $dc$ field
exceeds some threshold value at a given site of the crystal, which depends on
the state of the aging of the sample. The static domain structure gave no contrast.
Figure~\ref{fig3} illustrates the domain pattern evolution observed in SBN:70 crystal
sample at room temperature when external electric field is switched on. The
dark areas correspond to the areas which actually reversed their polarization
direction while clear areas are the regions where the switching process has not
started [initial polarization state in figure~\ref{fig3}~(a)] or has already been
completed. The switching process is realized by the nucleation and growth of
domains. We use the term ``nucleation'' to describe the emergence of new
antiparallel domains within the original domain as they appear in the video
image. However, it is an open question whether the initial domain state is
single-domain or contains the nano-scale residual domains not resolved by the
NLC method. It should be emphasized that the thickness of the visible ``wall''
is not a real physical thickness of the domain wall, which is much smaller (of the
order of several unit cells). The sidewise movement of the domain walls in
SBN:70 crystals is strongly perturbed by the random field environment related
to the relaxor properties and to the resulting domain wall pinning effect
\cite{Nattermann1990,Kleemann98}. Local structure disorder gives rise to quenched random
fields whose fluctuations are the source of local enhancements of the
coercivity. As a result of the pinning effect, some wall segments change their
position from image to image, whereas others stay immobile for a long time. As
a consequence, a maze type domain pattern is formed. A close inspection revealed
that individual domain walls expand with a great resistance under the $dc$ field and
finally clamp the entire dynamics. The pronounced slowing down of the domain
walls is reflected in a poor contrast of NLC above the slowly moving domain
walls, as can be seen in figure~\ref{fig3}~(d) and more clearly in figure~\ref{fig3}~(e), where
domain walls are hardly seen. The variations of the domain pattern are
noticeable following the application of the higher electric field of 1.3~kV/cm.
Simultaneously, a fine domain structure appears, as can be seen in figure~\ref{fig3}~(f).
With pinning centres present in the crystal, large domains are broken up into
smaller ones since certain areas of the domains are incapable of switching, and
consequently this process is accompanied by an increase of the domain density.
However, the polarization is fully reversed in the observed surface area after
a further poling in $E=1.5$~kV/cm. The switching process is completed first in the
central part of the image in figure~\ref{fig3}~(h) (where no obvious domain walls can be
identified), then in the outer region of the video scan. This field is also
sufficient to reverse the polarization state in the entire volume of the
examined SBN:70 crystal sample.

\begin{figure}[!tb]
\centerline{
\includegraphics[width=0.65\textwidth]{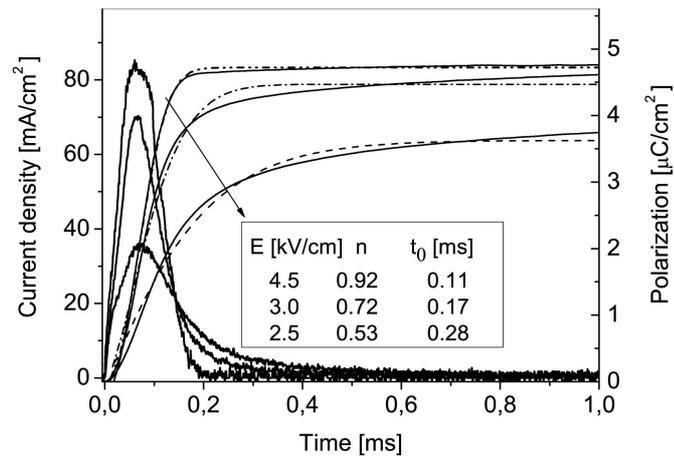}
}
\caption{The switching current and the switched polarization versus time for SBN:70 crystal sample. The broken curves correspond to the fit of the polarization data with a KWW function.
} \label{fig4}
\end{figure}

The fast polarization switching process was investigated by measuring the switching
currents in response to square wave electric pulses. A set of square pulses (of
10~ms duration) was applied by combining two positive and two negative pulses in
series. A true switching current was then obtained by subtracting the
nonswitching current from the switching one. Figure~\ref{fig4} presents the switching
curves obtained for various amplitudes of the electric fields for SBN:70 crystal
sample. The rate of polarization switching at constant $E$ can be found from
the switching current $i(t)$ by integration of $i(t)$ from  $t=0$ to the
instant $t$. The results are presented in figure~\ref{fig4}. The observed ``partial''
saturation of polarization may confirm the fact that there are slowly switching
regions that do not contribute to the switching current signals even in $E >
E_\mathrm{c}$. This should be related to the ``pinning effect'' of the domain walls in
SBN:70 crystal sample. The values of polarization obtained by the pulsed field
technique are comparable to that obtained from hysteresis-loop (H-L)
measurements at the same pulse amplitude of the $ac$-field at frequency 50~Hz.
The maximum value of polarization $\sim 8$~\SI{}{\micro\coulomb}/cm$^2$ found by H-L at
frequency $f=50$~Hz and $E=8$~kVcm$^{-1}$ is small compared to the total
polarization of about 20~\SI{}{\micro\coulomb}/cm$^2$ obtained from the measurements at a very
low-frequency of $E$ in which all domains can be aligned
\cite{Granzow01,Gladkii03}. It was found from pyroelectric measurements that a
total polarization of a polydomain crystal requires the field exposure times
that can range up to several tens of seconds even for the fields well above the
coercive field \cite{Granzow01}. In contrast to the stable H-L measured at high
frequency of $ac$-field, a non-coincidence of trajectories of H-L was obtained
during quasistatic loop registration in the course of a repeated field cycling
\cite{Granzow01,Gladkii03}. This process was accompanied by a considerable
decrease in the amplitude of polarization (the so-called fatigue effect).

The relaxation of polarization is well described by a Kohlrausch, Williams
and Watts (KWW) stretched exponential function $P(t)= P_0 \left\{1-\exp\left[-\left(t/t_0\right)^n\right]\right\}$
with $0 < n < 1$ widely used in dielectric relaxation studies. The fits are
represented by broken lines in figure~\ref{fig4}. The KWW function is usually used to
describe a complex relaxation and can  reflect a very broad distribution of
relaxation times \cite{Chamberlin98,Rogowski08}. This is closely related
to a distribution of random electric fields inherent to relaxor crystals. It is
interesting to note that KWW function described the switching kinetics in
SBN:61 \cite{Matyjasek12} as well as the relaxing domains on the nanoscale, with PFM
imaging of the domain configuration in SBN:61 doped with cerium \cite{Lehnen01}.

\subsection{Domain switching in SBN:26}

\begin{figure}[!t]
\centerline{
\includegraphics[width=0.65\textwidth]{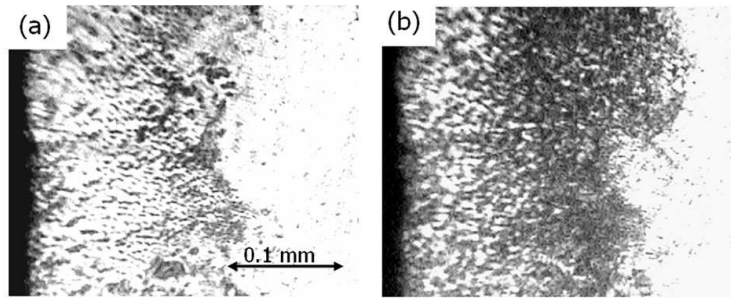}
}
\caption{Inhomogeneous distribution of domains observed on (001) plate of SBN:26 crystal sample in the electric field: (a) $+0.6$~kVcm$^{-1}$, (b) $-0.6$~kVcm$^{-1}$.} \label{fig5}
\end{figure}

\begin{figure}[!b]
\centerline{
\includegraphics[width=0.65\textwidth]{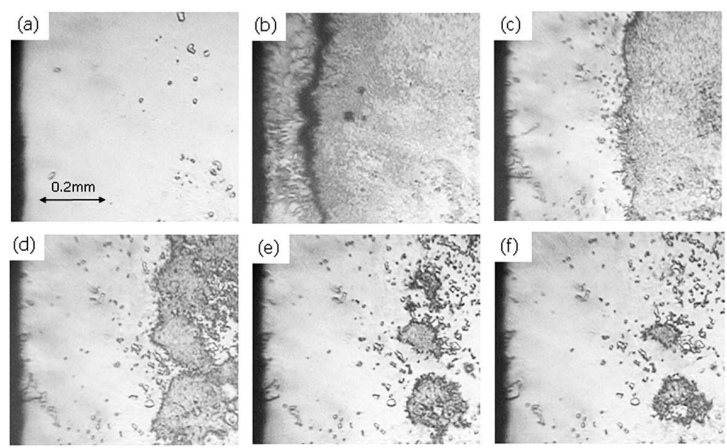}
}
\caption{Domain  pattern evolution observed in SBN:26 crystal sample in the electric field of 1~kVcm$^{-1}$. (a) initial single domain state. Time from the moment of applying $E$ (b) 0.4~s, (c) 4~s, (d) 6~s, (e) 7~s, (f) 8~s. Polarization axis is normal to the image plane.} \label{fig6}
\end{figure}

Microscopic observations of the domain pattern show that SBN:26 crystals usually
contain intrinsic defects in concentration high enough to effect the rate of
domain nucleation and growth in certain regions of the crystal sample. As is
shown in figure~\ref{fig5}, the distribution of nucleation sites is not random. Such
spatially non-uniform distribution of domain nuclei suggests the presence of a
frozen polarization component or built-in field, which favours one direction of
spontaneous polarization in certain regions. The fact that a similar picture
of delineated domains has been observed for a positive [figure~\ref{fig5}~(a)] and
a negative [figure~\ref{fig5}~(b)] polarization state, demonstrates the presence of a frozen
polarization state, possibly due to locally accumulated defects. The symmetry of
the hysteresis loop for SBN:26 crystal sample also indicates a lack of built-in
directional field, which could stabilize the domain structure in a preferential
direction. We observed a very slight variation in the domain size during the
switching process in low electric fields due to the domain wall pinning
effect. When the external field exceeds some threshold value (1~kV/cm for
the crystal sample examined), the domain walls begin to move from the crystal edge,
as is shown in figure~\ref{fig6}. The evolution of the domain pattern takes place via
expansion of the domain front formed after coalescence of newly created domains
at the edge of the crystal sample [figures~\ref{fig6}~(b)--(d)]. After a sufficient waiting
time ($\sim 0.4$~s) along with the progress in switching, the macroscopically visible
domains arise in the area before the moving domain front (in the region where
nucleation process is suppressed). The new domains are clearly seen in
figure~\ref{fig6}~(d). The domain wall front mobility is strongly inhomogeneous over the
crystal surface. A marked slowing down of the domain wall velocities was
observed during its propagation. Very approximate estimations give values from
$5\cdot10^{-3}$~m/s at the initial stage to $8\cdot10^{-6}$~m/s at the end
of the switching process in $E=1.0$~kV/cm. At higher electric fields, the
nucleated domains have been observed almost on the whole crystal surface, and
the switching speed increases.

\begin{figure}[!tb]
\centerline{
\includegraphics[width=0.5\textwidth]{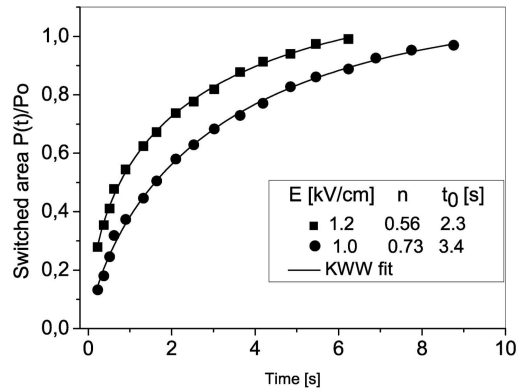}
}
\caption{Time dependence of the switched area for two different electric fields for domain configurations as shown in figure~\ref{fig6}.} \label{fig7}
\end{figure}

The pulse switching technique for studying the switching behaviour in higher
electric fields for SBN:26 crystals is inappropriate because the conduction
current obscures the displacement current. Kinetics of polarization relaxation
was determined in low electric fields by measuring the fraction of the
switched area as a function of time since the domain structure is naturally linked to
the polarization state of the crystal. Optical microscopy inspection on the
opposite polar faces showed a nearly complete penetration of the domains
throughout the crystal bulk. It means that the forward domain growth along the
polarization direction takes place very quickly, and the sidewise domain growth
determines the kinetics of polarization switching. Thus, the domain growth becomes a two-dimensional problem. Therefore, the area of the
switched domains normalized to the total scan area is expected to be close to
the normalized reverse polarization. The recording of the domain patterns was performed with a digital camera. In order to obtain a quantitative description of the polarization evolution, the data have been processed using an image analysis program. Figure~\ref{fig7} exemplarily illustrates  the
evolution of the switched areas as a function of time under two different
applied fields. The data points are well fitted (solid lines) with KWW function,
where the values of the stretching exponent, $n$, are given in figure~\ref{fig7}.

\section{Discussion}

It is interesting to note that the domain switching process in SBN:70 single
crystals is similar to that observed in Ni doped SNB:61 single crystals
\cite{Matyjasek12}. It was shown that Ni doping additionally deteriorates
the homogeneity of the
SBN:61 crystals. In both type of crystals, only the electric field
$E>E_\mathrm{c}$ is sufficient to complete the microscale domain switching in the entire
volume of the crystal sample. In the doped SBN:61 and SBN:70 single crystals, slow
and fast domain walls coexist, which could play an important role in the
ferroelectric phase transition broadening, assuming the formation of polar
regions with locally different Curie temperature. On the other hand, the domain
switching in SBN:26 crystal in the high-field regime may qualitatively resemble
the switching observed in classical ferroelectric crystals, in which  the growth of
the existing domains is more favourable than the creation of new ones due to
a relatively low domain-wall energy \cite{Miller60}. Thus, the nucleated
domains expand with no or little resistance under an electric field.  The SBN:26
crystal exhibits a high electrical conductivity. There has been reported a
considerable increase of the leakage current density since the applied field exceeds a
critical value in SBN crystals with low Sr content \cite{Qu02,Matyjasek07}.
Once the domains are nucleated in the high-field regime, the field induced
charge carriers may compensate the sources of random fields giving rise to the
depinning of the domain walls in SBN:26 crystals. It has been reported that having
raised the sample conductivity by illumination, the pinning centres were eliminated
and the polarization dynamics was fully restored in doped SBN:61 crystals
\cite{Granzow02}.

The relaxation of polarization during the switching process can be described
in the examined single crystals by KWW stretched exponential function. It is
interesting to note that our results are consistent with a general concept saying
that dynamical heterogeneity was established as a possible source of a stretched
exponential relaxation \cite{Chamberlin98}.

The domain walls make a considerable contribution  to the total dielectric
response of the SBN system investigated. One can presume that a weak measuring
$ac$ field in dielectric measurements does not affect the domain configuration
but creates nuclei on the domain walls. Our results may confirm that the
composition disorder could play an important role in the dielectric response in
SBN:70. A broad distribution of the heights of local pinning barriers should
yield a distribution of activation energies for the nucleation of reversed steps
on the existing domain walls. Consequently, SBN:70 crystals show much
broader temperature dependence and lower values of dielectric susceptibility
than those observed in SBN:26 crystals, in which the motion of domain walls is
less restricted by a random field environment. Dielectric measurements as well as
the domain switching observations have evidenced the change from typical relaxor
behaviour in SBN:70 into classic ferroelectric behaviour in the case of SBN:26
crystal. It is interesting to note that the crossover from  relaxor to
ferroelectric behaviour was also observed in solid solutions of
BaTi$_{1-x}$Sn$_x$O$_3$, and was evidenced by dielectric spectroscopic data
\cite{Shvartsman08b}.

It must be noted that visualization of the domain structure by NLC method may
not thoroughly reflect the domain structure evolution in electrical switching at
the same pulse amplitudes. Due to a slow response time (of the order of
several ms), liquid crystal molecules may not follow the changes of the domain
structure in the regions that exhibit high speed switching. Moreover, the results of
the works show a strong effect of an interface (electrode and ferroelectric
surface) conductivity on the kinetics of the switching process
\cite{Shur06}. In a metal contact, electrical charges can freely move and
can effectively screen the depolarization field accompanying the ferroelectric
polarization, which decreases the switching process. However, visualization of
a domain structure by NLC-method can be used in checking the quality of the
crystal for the purpose of choosing high quality SBN samples with a homogeneous distribution of domain nuclei.

\section{Conclusions}

The NLC decoration  technique is a relatively simple method to visualize the
distribution and evolution of the microscale domains during the switching
process, which obviously reflects the internal disorder of a crystal. The
domain structure dynamics confirm the dielectric measurements stating that the
ferroelectric properties of SBN system change with the composition from strong
relaxor behaviour in SBN:70 into classical ferroelectrics as observed in the
case of SBN:26. Characteristic features of the domain structure dynamics in SBN
crystal of different composition (SBN:70 and SBN:26) can serve as a direct
verification of the concept of local random fields that exist due to a structural
disorder of the SBN crystal. Fluctuations of random fields neither effect
the entire domains nor form a macroscopic bias field, but act as pinning centres for
the domain walls in SBN:70 crystal.  For SBN:26 crystal, the domain walls can
only be pinned in low electric fields. At higher fields, the field- induced
charge carriers may compensate the sources of random fields giving rise to a
less restricted domain growth. The polarization switching process in both
crystals can be described by a stretched exponential function widely used in
dielectric relaxation studies.

\section*{Acknowledgements}

Thanks are due to the Polish Ministry of Science for a partial financial support
under grant \linebreak No.~N~N507~455034. Professor W.~Kleemann from the University of
Duisburg-Essen is acknowledged for making accessible the Solartron Impedance
Analyzer.


\begin{thebibliography}{99}
\bibitem{Ramirez12} Ramirez~M.O., Molina~P.,  Baus\'{a}~L.E.,  Opt. Mater., 2012,
\textbf{34}, 524; \doi{10.1016/j.optmat.2011.03.016}.

\bibitem{Soergel05} Soergel~E., Appl. Phys. B, 2005, \textbf{81}, 729; \doi{ 10.1007/s00340-005-1989-9}.

\bibitem{Shur11} Shur~V.Ya., Shikhova~V.A., Pelegov~D.V., Ievlev~A.V., Ivleva~L.
I., Phys. Solid State, 2011, \textbf{53}, 2311; \\ \doi{10.1134/S106378341111028X}.

\bibitem{Terabe02} Terabe~K., Takekawa~S., Nakamura~M., Kitamura~K.,
Higuchi~S., Gotoh~Y., Gruverman~A., Appl. Phys. Lett., 2002,
\textbf{81}, 2044; \doi{10.1063/1.1506945}.

\bibitem{Lehnen01} Lehnen~P., Kleemann~W., Woike~Th., Pankrath~R., Phys. Rev.
B, 2001, \textbf{64}, 224109; \doi{10.1103/PhysRevB.64.224109}.

\bibitem{Shvartsman08} Shvartsman~V.V., Kleemann~W., \L{}ukasiewicz~T.,
Dec~J., Phys. Rev. B, 2008, \textbf{77}, 054105; \\ \doi{10.1103/PhysRevB.77.054105}.

\bibitem{Liu09} Liu~X.Y., Liu~Y.M., Takekawa~S., Kitamura~K., Ohuchi~F.S.,
Li~J.Y., J. Appl. Phys., 2009, \textbf{106}, 124106; \\ \doi{10.1063/1.3273481}.

\bibitem{Gainutdinov09} Gainutdinov~R.V., Volk~T.R., Lysova~O.A., Razgonov~I.I.,
Tolstikhina~A.L., Ivleva~L.I., Appl. Phys. B, 2009, \textbf{95}, 505; \doi{10.1007/s00340-009-3507-y}.

\bibitem{Tian05} Tian~L., Scrymgeour~D.A., Gopalan~V., J. Appl. Phys.,
2005, \textbf{97}, 114111; \doi{10.1063/1.1925330}.

\bibitem{Shur08} Shur~V.Ya., Pelegov~D.V., Shikhova~V.A., Kuznetsov~D.K.,
Nikolaeva~E.V., Rumyantsev~E.L., Yakutova~O.V., Granzow~T., Ferroelectrics,
2008, \textbf{374}, 33; \doi{10.1080/00150190802424785}.

\bibitem{Ivanov02}  Ivanov~N.R., Volk~T.R., Ivleva~L.I., Chumakova~S.P.,
Ginsberg~A.V., Cryst. Rep., 2002, \textbf{47}, 1023; \\ \doi{10.1134/1.1523521}.

\bibitem{Matyjasek08} Matyjasek~K., Wolska~K., Kaczmarek~S.M.,
Rogowski~R.Z., J. Phys.: Condens. Matter, 2008, \textbf{20}, 295218; \\ \doi{10.1088/0953-8984/20/29/295218}.

\bibitem{Matyjasek12} Matyjasek~K., Wolska~K., Kaczmarek~S.M., Subocz~J.,
 Ivleva~L.I., Appl. Phys. B, 2012, \textbf{106}, 143; \\ \doi{10.1007/s00340-011-4773-z}.

\bibitem{Jamieson68} Jamieson~P.B., Abrahams~S.C., Bernstein~J.L., J. Chem.
Phys., 1968, \textbf{48}, 5048; \doi{10.1063/1.1668176}.

\bibitem{Nattermann1990} Nattermann~T., Shapir~Y., Vilfan~I., Phys.Rev. B, 1990, \textbf{42}, 8577; \doi{DOI:10.1103/PhysRevB.42.8577}.

\bibitem{Kleemann98} Kleemann~W., Phase Trans., 1998, \textbf{65}, 141; \doi{10.1080/01411599808209285}.

\bibitem{Qu02} Qu~Y.Q., Li~A.D., Shao~Q.Y., Tang~Y.F., Wu~D.,  Mak~C.L.,
Wong~K.H., Ming~N.B., Mater. Res. Bulletin, 2002, \textbf{37}, 503; \doi{10.1016/S0025-5408(02)00676-1}.

\bibitem{Lukasiewicz08} \L{}ukasiewicz~T., Swirkowicz~M.A., Dec~J., Hofman~W.,
Szymski~W.J., J. Cryst. Growth, 2008, \textbf{310}, 1464; \\ \doi{10.1016/j.jcrysgro.2007.11.233}.

\bibitem{Santos09} Santos~I.A., Mendes~R.G., Eiras~J.A., de~Los~J., Guerra~S.,
Ara\'{u}jo~E.B., Appl. Phys. A, 2009, \textbf{95}, 757; \\ \doi{10.1007/s00339-008-5060-7}.

\bibitem{Tikhomirova79} Tikhomirova~N.A., Dontsova~L.J., Pikin~S.A.,
Shuvalov~L.A., JETP  Lett., 1979, \textbf{29}, 34.

\bibitem{Granzow01} Granzow~T., Doerfler~U., Woike~Th., Woehlecke~M., Pankrath~R.,
Imlau~M., Kleemann~W., Phys. Rev. B, 2001, \textbf{63}, 174101; \doi{10.1103/PhysRevB.63.174101}.

\bibitem{Gladkii03} Gladkii~V.V., Kirikov~V.A., Volk~T.R.,  Isakov~D.V.,
Ivanova~E.S., Phys. Solid State, 2003, \textbf{45}, 2171; \\ \doi{10.1134/1.1626758}.

\bibitem{Chamberlin98} Chamberlin~R.V., Phase Transit., 1998,
\textbf{65}, 169; \doi{10.1080/01411599808209287}.

\bibitem{Rogowski08} Rogowski~R.Z., Matyjasek~K., Wolska~K., Kaczmarek~S.M.,
Phase Transit., 2008, \textbf{81}, 1039; \\ \doi{10.1080/01411590802457946}.

\bibitem{Miller60} Miller~R.C., Weinreich~G., Phys. Rev., 1960, \textbf{117},
1460; \doi{10.1103/PhysRev.117.1460}.

\bibitem{Matyjasek07} Matyjasek~K., Repow~K., Kaczmarek~S.M., Berkowski~M., J.~Phys.: Condens. Matter, 2007, \textbf{19}, 466207; \\ \doi{10.1088/0953-8984/19/46/466207}.

\bibitem{Granzow02} Granzow~T., Doerfler~U., Woike~T., Woehlecke~M.,
Pankrath~R., Imlau~M., Kleemann~W., Europhys. Lett., 2002, \textbf{57}, 597; \doi{10.1209/epl/i2002-00503-6}.

\bibitem{Shvartsman08b} Shvartsman~V.V., Dec~J., Xu~Z.K., Banys~J., Keburis~P.,
Kleemann~W., Phase Transit., 2008, \textbf{81}, 1013; \\ \doi{10.1080/01411590802457888}.

\bibitem{Shur06}  Shur~V.Ya., J. Mater. Sci., 2006, \textbf{41}, 199; \doi{10.1007/s10853-005-6065-7}.


\end{thebibliography}

\newpage
\ukrainianpart

\title{Сегнетоелектричні і діелектричні дослідження стронцій-барієвих ніобатів релаксорного і
сегнетоелектричного типу}

\author {К. Матиясек\refaddr{label1},
        Я. Дец \refaddr{label2},
          С. Міґа\refaddr{label2},
          Т. Лукасєвіч\refaddr{label3}    }

\addresses{
\addr{label1} Інститут фізики, факультет механічної інженерії і
мехатроніки, \\
Західно-померанський технологічний університет, Щецін, Польща
\addr{label2} Інститут матеріалознавства, Сілезький університет,
Катовіце, Польща %
\addr{label3} Інститут технології матеріалів електроніки, Варшава, Польща }

\makeukrtitle
\begin{abstract}

Еволюція сегнетоелектричної доменної структури, індукованої
зовнішнім електричним полем досліджувалась за допомогою методу
нематичного рідкого кристалу  в двох монокристалах
стронцій-барієвого ніобату номінального складу:
Sr$_{0.70}$Ba$_{0.30}$Nb$_{2}$O$_{6}$ (SBN:70~--- релаксорний) і
Sr$_{0.26}$Ba$_{0.74}$Nb$_{2}$O$_{6}$ (SBN:26~---
сегнетоелектричний). Наші результати показують, що широкий фазовий
перехід і частотна дисперсія,  продемонстровані  кристалом SBN:70,
мають тісний зв'язок із конфігурацією сегнетоелектричних
мікродоменів. Великий струм стікання,  виявлений в SBN:26, може
компенсувати внутрішні заряди, що діють як центри пінінгу для
доменних стінок, що приводить до менш обмеженого  росту доменів
подібно до того, що спостерігається в класичних сегнетоелектриках.
Мікромасштібні дослідження процесу перемикання в поєднанні з
електричними вимірюваннями дозволяють встановити співвідношення між
локальними властивостями динаміки доменів і макроскопічним відгуком,
а саме, гістерезисною петлею поляризації і діелектричними
властивостями.
\keywords релаксорний сегнетоелектрик, ніобати, доменні стінки,
процес перемикання, SBN
\end{abstract}

\end{document}